\documentstyle[12pt,aaspp4]{article}

\begin{document}

\title{Supernovae as the Site of the $r$-Process: Implications for
Gamma-Ray Astronomy}
\author{Y.-Z. Qian and P. Vogel}
\affil{Department of Physics, California Institute of Technology,
       Pasadena, CA 91125}
\authoremail{yzqian@citnp.caltech.edu, vogel@lamppost.caltech.edu}
\and
\author{G. J. Wasserburg}
\affil{The Lunatic Asylum,
Division of Geological and Planetary Sciences, California
       Institute of Technology, Pasadena, CA 91125}

\begin{abstract}

We discuss how
detection of gamma-ray emission from the decay of $r$-process nuclei
can improve our understanding of $r$-process nucleosynthesis.
We find that a gamma-ray detector
with a sensitivity of $\sim 10^{-7}~\gamma$~cm$^{-2}$~s$^{-1}$
at $E_\gamma\approx 100$--700 keV may detect
the emission from the decay of $^{125}$Sb, $^{137}$Cs, $^{144}$Ce,
$^{155}$Eu, and $^{194}$Os produced in a future Galactic supernova.
In addition, such a detector may detect
the emission from the decay of $^{126}$Sn in the Vela supernova remnant
and the diffuse emission from the decay of $^{126}$Sn produced by 
past supernovae in our Galaxy.
The required detector sensitivity is similar to what is projected for
the proposed Advanced Telescope for High Energy Nuclear Astrophysics
(ATHENA). Both the detection of gamma-ray emission from the
decay of several $r$-process nuclei 
(e.g., $^{125}$Sb and $^{194}$Os) produced in future
Galactic supernovae and the detection of emission 
from the decay of $^{126}$Sn in the
Vela supernova remnant would prove that supernovae are a site of 
the $r$-process. Furthermore, the former detection would
allow us to determine whether or not the $r$-process nuclei
are produced in relative proportions specified by
the solar $r$-process abundance pattern in supernova $r$-process events.
Finally, detection of 
diffuse emission from the decay of $^{126}$Sn in our Galaxy would eliminate
neutron-star--neutron-star mergers as the main source for the $r$-process
nuclei near mass number $A\sim 126$.

\end{abstract}


\section{Introduction}

Approximately half the natural abundance 
of heavy elements with mass number $A>70$ and all of the actinides
in the solar system came from the $r$-process. Although the
$r$-process theory was put forward by Burbidge et al. (1957) and
Cameron (1957) more than four decades ago, the site of the $r$-process
has remained a mystery. The extreme conditions (e.g., neutron number
densities exceeding $\sim 10^{20}$~cm$^{-3}$ and timescales of
$\sim 1$~s) required for the $r$-process suggest that it might
occur in some violent astrophysical events such as core-collapse
supernovae (i.e., Type II and Type Ib supernovae, hereafter referred to simply
as supernovae). 
A similarly violent but more exotic site
of the $r$-process might be neutron-star--neutron-star (NS-NS) mergers
(see Cowan, Thielemann, \& Truran 1991 for a review of the possible
$r$-process sites).

Another question closely related to the site of the $r$-process 
concerns the relative production of $r$-process nuclei in individual
events. Observation of $r$-process elements in extremely
metal-poor halo stars has shed some important light on this question.
For example, Sneden et al. (1996) found that
the $r$-process abundance pattern at and beyond Ba (i.e., at $A\ge 135$)
in CS22892--052 is consistent with that in the solar system.
This suggests that the solar $r$-process abundance pattern, especially
the part at $A\ge 135$, may be
generic to the $r$-process, and thus may characterize the production
in every event. However, elements in the solar $r$-process
abundance peak at $A\sim 130$ have not yet been detected in CS22892--052.
Furthermore, since it is difficult to establish
that the $r$-process elements in this star were produced in a single
event, the observed $r$-process abundance pattern could
still be a superposition of those from (intrinsically) different events.
Therefore, the relevance of the solar $r$-process
abundance pattern to the production in individual events
remains to be established.

In this paper we discuss how detection of
gamma-ray emission from the decay of $r$-process nuclei may
answer the following questions: (1) Are supernovae a site
of the $r$-process? (2) Are the $r$-process nuclei produced
in relative proportions specified by the solar $r$-process abundance
pattern in supernova $r$-process events? (3) Are NS-NS mergers
a main source for some $r$-process nuclei?

Gamma-ray emission characteristic of a radioactive 
nucleus from an astrophysical event would provide direct evidence for
production of this nucleus in this event. 
Clayton, Colgate, \& Fishman (1969) predicted that such emission
might be detectable. More recently, Meyer \& Howard (1991) 
discussed possible gamma-ray signatures of an $r$-process event. 
In particular, with consideration of supernovae as the $r$-process site,
they estimated gamma-ray fluxes from the decay of some
$r$-process nuclei produced in SN 1987A. They also discussed
the possibility of detecting such fluxes.

In the present work, we generalize the approach of Meyer \& Howard (1991)
with particular consideration of future gamma-ray detectors. We find that
if supernovae are the site of the $r$-process, a number of 
$r$-process nuclei, namely $^{125}$Sb, $^{137}$Cs, $^{144}$Ce,
$^{155}$Eu, and $^{194}$Os,
can provide gamma-ray fluxes of $\sim 10^{-7}~\gamma$~cm$^{-2}$~s$^{-1}$
from a future Galactic supernova at a distance of 10 kpc.
In addition, we show that the decay of $^{126}$Sn in the Vela supernova remnant
can produce fluxes of a similar magnitude. 
Detection of such fluxes would be possible
for a gamma-ray detector with a sensitivity similar to
what is projected for the proposed
Advanced Telescope for High Energy Nuclear Astrophysics
(ATHENA) (Kurfess 1994).
Such a detector may also detect the diffuse
gamma-ray emission from the decay of $^{126}$Sn produced
by past supernovae in the Galaxy.
Both the detection of gamma-ray emission from the
decay of several $r$-process nuclei 
(e.g., $^{125}$Sb and $^{194}$Os) produced in future Galactic
supernovae and the detection of emission from the decay of $^{126}$Sn
in the Vela supernova remnant would prove that supernovae are
a site of the $r$-process, thus answering the first question posed above. 
In addition, the former detection would
allow us to determine whether or not
the $r$-process nuclei are produced in relative
proportions specified by the solar $r$-process abundance pattern in
supernova $r$-process events, thus answering the second question. 
The third question would be answered by detection of diffuse
gamma-ray emission from the decay of $^{126}$Sn in the Galaxy, since this
detection would eliminate NS-NS mergers
as the main source for the $r$-process nuclei near $A\sim 126$.

The organization of this paper is as follows. In \S2 we estimate 
gamma-ray fluxes from the decay of a number of $r$-process nuclei
(e.g., $^{125}$Sb and $^{194}$Os)
produced in a future Galactic supernova and from the decay of $^{126}$Sn
in the Vela supernova remnant. Diffuse fluxes from the decay of
$^{126}$Sn produced by past supernovae in the Galaxy are also estimated.
In \S3 we discuss the results of \S2 in connection with answers to the
three questions posed above. We give our conclusions in \S4.

\section{Gamma-Ray Emission from the Decay of $r$-Process Nuclei}

We assume that supernovae are the site of the
$r$-process in this section.
Gamma-ray emission from the decay of
nuclei produced in a supernova may be detected only
after it becomes transparent to gamma rays. In the case of
Type II supernovae, the delay between the explosion and
onset of gamma-ray transparency 
is typically several years. This delay is shorter if
strong mixing occurs in the envelope as for SN 1987A
(e.g., Arnett et al. 1989), or if
the supernova progenitor does not have a hydrogen envelope (e.g.,
Type Ib supernovae). Unambiguously, $r$-process nuclei that may be
of interest to gamma-ray astronomy must have lifetimes of 
$\sim 1$~yr or longer.
A search through the Table of Isotopes (Firestone et al. 1996) identifies
seven such nuclei. The five relatively short-lived ones are $^{125}$Sb,
$^{137}$Cs, $^{144}$Ce, $^{155}$Eu, and $^{194}$Os (see Table 1),
and the two long-lived
ones are $^{126}$Sn (see Table 2) and $^{182}$Hf. 
We emphasize that all of these
nuclei are bypassed by the $s$-process, and 
are made only in the $r$-process. 

\subsection{The Relatively Short-Lived $r$-Process Nuclei}

Lifetimes of the five relatively short-lived $r$-process nuclei range from
1.12 yr for $^{144}$Ce to 43.4 yr for $^{137}$Cs (see Table 1). 
Since these lifetimes
are shorter than or comparable to the average interval 
($\sim 30$~yr) between
successive supernovae in the Galaxy,
one has to wait for a future Galactic supernova to detect
the gamma-ray emission from the decay of these nuclei.

These five relatively short-lived $r$-process nuclei were also
identified by Meyer \& Howard (1991), as well as 
$^{90}$Sr, $^{106}$Ru, $^{151}$Sm, and $^{171}$Tm.
We do not include $^{90}$Sr, $^{151}$Sm, and $^{171}$Tm 
in our discussion since
fluxes from their decay are much lower than those
from the decay of the other relatively short-lived $r$-process nuclei. 
Furthermore, $^{90}$Sr and $^{106}$Ru are commonly produced in the
$\alpha$-process (Woosley \& Hoffman 1992), and thus cannot provide 
signatures unique to the $r$-process. [In fact, observation by Sneden et al.
(1998) has shown that abundances of Sr, Y, and Zr (with $A\sim 90$)
in very metal-poor halo stars do not fit the solar $r$-process abundance
pattern whereas those of the elements including and beyond Ba do.]
Therefore, detection of gamma-ray emission from the 
decay of $^{106}$Ru 
would not directly improve our understanding of $r$-process nucleosynthesis
in contrast to the case of pure $r$-process nuclei.

Assuming that the solar
$r$-process composition represents the Galactic average, we can
estimate the average amount of mass in a radioactive $r$-process nucleus 
produced in a supernova as
\begin{equation}
\delta M\approx{X_{\odot}^rM_{G}\over f_{\rm SN}t_{G}}
=3\times 10^{-7}\,M_\odot\left({X_{\odot}^r\over 10^{-9}}\right)
\left({M_{G}\over 10^{11}\,M_\odot}\right)
\left({10^{10}\ {\rm yr}\over t_{G}}\right)
\left({f_{\rm SN}^{-1}\over 30\ {\rm yr}}\right),
\label{mass}
\end{equation}
where $X_{\odot}^r$ is the solar $r$-process mass fraction of the stable
decay product of this nucleus (e.g., for $^{125}$Sb, $X_{\odot}^r$ is
from $^{125}$Te),
$M_{G}$ and $t_{G}$ are the total mass and the age of the Galaxy,
respectively, and $f_{\rm SN}$ is the frequency within the Galaxy
for the supernovae
which produce this nucleus. We calculate $X_{\odot}^r$ in
equation (\ref{mass}) for any specific nucleus from its solar
mass fraction given by Arnett (1996) and the corresponding solar
$r$-process fraction given by K\"appeler, Beer, \& Wisshak (1989).
Subtraction of the $s$-process contribution in deriving the solar
$r$-process fraction gives rise to a typical relative
uncertainty of $\sim\pm 10\%$
in $X_{\odot}^r$. An exception is $^{137}$Ba, with a relative uncertainty
of $\sim\pm 100\%$ in $X_{\odot}^r$
(i.e., in the extreme case, there is no $r$-process production of
the corresponding radioactive progenitor $^{137}$Cs).
As discussed in \S3, the largest
uncertainty in $\delta M$ comes from the intrinsic variation 
of the amount of $r$-process
production in individual supernovae. So the value of $\delta M$
in equation (\ref{mass}) should be taken as a rough
estimate for production in a specific supernova.

If we assume that the delay between 
production of the $r$-process nucleus and 
onset of gamma-ray transparency of the supernova
is small compared with the lifetime of this nucleus,
then the gamma-ray flux to be detected from its decay is
\begin{equation}
F_\gamma={N_A\over 4\pi d^2}{\delta M\over A}{I_\gamma\over\bar\tau}
=3.2\times 10^{-6}\,I_\gamma\left({\delta M\over 10^{-7}\,M_\odot}\right)
\left({100\over A}\right)\left({1\ {\rm yr}\over\bar\tau}\right)
\left({10\ {\rm kpc}\over d}\right)^2\ \gamma~{\rm cm}^{-2}~{\rm s}^{-1},
\label{flux}
\end{equation}
where $d$ is the distance to the supernova, 
$N_A$ is Avogadro's number, $A$ is the mass number of this
nucleus, $\bar\tau$ is its lifetime, and $I_\gamma$ is the number of photons
emitted at a specific energy $E_\gamma$ per decay of this nucleus.
The expected gamma-ray
fluxes from the decay of the five relatively short-lived $r$-process
nuclei are given in Table 1 for a supernova 
at a distance of 10 kpc, 
along with the expected mass $\delta M$ in the corresponding
nuclei produced in the supernova. These fluxes differ somewhat from the
results (scaled to a distance of 10 kpc) of Meyer \& Howard (1991) as
they used somewhat different estimates of the $r$-process
production $\delta M$ and apparently, 
different nuclear parameters in some cases.

\subsection{The Long-Lived $r$-Process Nuclei}

The nucleus $^{126}$Sn ($\bar\tau=1.44\times 10^5$~yr)
can decay to both the ground and the
first excited states of $^{126}$Sb, which in turn decay to 
the stable $^{126}$Te
with lifetimes of 18 days and 27.6 minutes, respectively.
Three prominent gamma rays at $E_\gamma=415$, 666,
and 695 keV are emitted with $I_\gamma\approx 1$ in the decay chain
$^{126}{\rm Sn}\rightarrow{^{126}{\rm Sb}}\rightarrow{^{126}{\rm Te}}$.
Due to the long lifetime of $^{126}$Sn, substantial
gamma-ray fluxes are expected only from a nearby supernova
(cf. eq. [\ref{flux}]). If we choose a reference value of $d=200$~pc
for the distance to the supernova, the corresponding flux from the
decay of $^{126}$Sn is
\begin{equation}
F_\gamma\approx 2.2\times 10^{-7}I_\gamma
\left({\delta M\over 5\times 10^{-7}\,M_\odot}\right)
\left({200\ {\rm pc}\over d}\right)^2\ \gamma~{\rm cm}^{-2}~{\rm s}^{-1}.
\label{vela}
\end{equation}
In equation (\ref{vela}),
the reference value of $\delta M=5\times 10^{-7}\,M_\odot$ for the
amount of $^{126}$Sn production is estimated
from equation (\ref{mass}) with
$X_{\odot}^r({^{126}{\rm Te}})=1.65\times 10^{-9}$,
$M_{G}=10^{11}\,M_\odot$,
$t_{G}=10^{10}$ yr, and $f_{\rm SN}=(30\ {\rm yr})^{-1}$.

Interestingly, the distance to the Vela pulsar is estimated to
be only about 125--500 pc (Milne 1968; Oberlack et al. 1994; 
Aschenbach, Egger, \& Tr\"umper 1995; Becker 1995). 
Furthermore, its age is estimated to be about $10^4$ yr,
much less than the lifetime of $^{126}$Sn. So if the supernova associated
with the Vela pulsar produced $^{126}$Sn, then most of
the radioactive $^{126}$Sn nuclei initially produced in the supernova
will remain there for a very long time. The expected 
gamma-ray fluxes from the decay of
$^{126}$Sn in the Vela supernova remnant are given in Table 2 for
$d=200$ pc (cf. Oberlack et al. 1994) 
and $\delta M=5\times 10^{-7}\,M_\odot$. 
[We note that gamma-ray emission
from the decay of $^{26}$Al has been detected from the Vela region
(Diehl et al. 1995). In addition, an optical search for 
some $r$-process elements in the Vela supernova remnant was
carried out by Wallerstein et al. (1995).]

For an $r$-process nucleus such as $^{126}$Sn
with $f_{\rm SN}^{-1}\ll\bar\tau\ll t_G$, 
production by past supernovae that occurred over a long timescale
can provide a substantial abundance of this nucleus in the Galaxy,
and therefore can give rise to diffuse gamma-ray emission.
By ``diffuse,'' we mean that the emission comes from a collection of
unresolved point sources. Under the assumption that
such a nucleus was uniformly produced over the Galactic history, its
present abundance is
\begin{equation}
N\approx N_A{p\bar\tau\over A},
\end{equation}
where
\begin{equation}
p\approx{X_{\odot}^rM_G\over t_G}=10^{-8}
\left({X_{\odot}^r\over 10^{-9}}\right)
\left({M_{G}\over 10^{11}\,M_\odot}\right)
\left({10^{10}\ {\rm yr}\over t_{G}}\right)M_\odot\ {\rm yr}^{-1}
\label{up}
\end{equation}
is its rate of production by mass in the Galaxy. As we have taken
$f_{\rm SN}^{-1}\ll\bar\tau$, there would have been 
a large number of supernova 
contributions to the diffuse emission, which can 
be distributed quite irregularly through the Galaxy.
In fact, diffuse emission from the decay of $^{26}$Al
predicted by Ramaty \& Lingenfelter (1977) has been observed 
to be irregular (see
Prantzos \& Diehl 1996 for a review).
If the Galactic distribution of a radioactive $r$-process nucleus is roughly
the same as that of interstellar hydrogen (such a distribution is also
consistent with that deduced for supernova remnants and pulsars), then 
according to Mahoney et al. (1982), the corresponding average
diffuse flux 
at longitudes within $\pm 30^\circ$ of the Galactic center is
\begin{equation}
{\cal{F}}_\gamma\approx 1.0\times 10^{-46}N_A{p\over A}I_\gamma
=3.8\times 10^{-7}I_\gamma
\left({p\over 10^{-8}\,M_\odot~{\rm yr}^{-1}}\right)
\left({100\over A}\right)
\ \gamma\ {\rm cm}^{-2}~{\rm s}^{-1}~{\rm rad}^{-1}.
\label{diff0}
\end{equation}
The ``rad'' in the unit for ${\cal{F}}_\gamma$
refers to the Galactic longitude.
Note that ${\cal{F}}_\gamma$ is independent of $\bar\tau$ for
$f_{\rm SN}^{-1}\ll\bar\tau\ll t_G$.

From equation (\ref{diff0}), 
the average diffuse flux from the decay
of $^{126}$Sn in the Galactic center direction is
\begin{equation}
{\cal{F}}_\gamma\approx 5.0\times 10^{-7}I_\gamma
\left({p\over 1.65\times 10^{-8}\,M_\odot~{\rm yr}^{-1}}\right)
\ \gamma~{\rm cm}^{-2}~{\rm s}^{-1}~{\rm rad}^{-1}.
\label{diff}
\end{equation}
In equation (\ref{diff}), the reference value of 
$p=1.65\times 10^{-8}\,M_\odot~{\rm yr}^{-1}$ is estimated from
equation (\ref{up}) with $X_\odot^r({^{126}{\rm Te}})=1.65\times 10^{-9}$,
$M_G=10^{11}\,M_\odot$, and $t_G=10^{10}$ yr. The diffuse fluxes 
from the decay of $^{126}$Sn are also
given in Table 2.

Similar calculations for $^{182}$Hf
[$\bar\tau=1.3\times 10^7$ yr and
$X_\odot^r({^{182}{\rm W}})=5.3\times 10^{-11}$] show that
although many gamma rays are emitted in the decay chain
${^{182}{\rm Hf}}\rightarrow{^{182}{\rm Ta}}\rightarrow{^{182}{\rm W}}$,
the flux from the Vela supernova remnant is only 
$F_\gamma\sim 4.3\times 10^{-11}\ \gamma~{\rm cm}^{-2}~{\rm s}^{-1}$ and
the diffuse flux in the Galactic center direction is only
${\cal{F}}_\gamma
\sim 8.9\times 10^{-9}\ \gamma~{\rm cm}^{-2}~{\rm s}^{-1}~{\rm rad}^{-1}$
for the most prominent line at $E_\gamma=270$ keV with $I_\gamma=0.8$.

\section{Discussion}

From the results in Tables 1 and 2,
we can see that a gamma-ray detector with a sensitivity of
$\sim 10^{-7}\ \gamma~{\rm cm}^{-2}~{\rm s}^{-1}$ at
$E_\gamma\approx 100$--700 keV may detect (1) 
fluxes from the decay of
a number of $r$-process nuclei such as $^{125}$Sb
and $^{194}$Os produced in a future Galactic supernova, (2)
fluxes from the decay of $^{126}$Sn
in the Vela supernova remnant, and (3)
diffuse fluxes from the decay of $^{126}$Sn produced by 
past supernovae in the Galaxy. Each of these detection
possibilities and its significance for our understanding of $r$-process 
nucleosynthesis are discussed below.

\subsection{Fluxes from a Future Galactic Supernova}

A future Galactic supernova will announce itself by a powerful neutrino
burst even if its optical display is obscured from us. In addition,
forward-peaked neutrino-electron scattering events in a water
\v Cerenkov detector such as Super-Kamiokande can provide directional
information about the supernova. Even the distance to the supernova can 
be estimated from the total number and the energies of 
detected neutrino events
since the total energy emitted in neutrinos is known within a factor of
a few. The delay between
the explosion and onset of gamma-ray transparency leaves time for directing
suitable detectors to search for gamma-ray emission
from the decay of the $r$-process nuclei listed
in Table 1. Except possibly for $^{144}$Ce, these nuclei will have
most of their initial abundance produced in the supernova when it
becomes transparent. If gamma-ray emission
from the decay of the $r$-process nuclei listed in Table 1 were detected
from the supernova, this would prove that these nuclei are produced
in supernovae. 

In addition, we are interested in determining
whether or not these nuclei are produced
in relative proportions specified by the solar $r$-process
abundance pattern in supernova $r$-process events.
If every supernova $r$-process event were to produce 
the same $r$-process abundance pattern as that in the solar system, 
gamma-ray
fluxes from the decay of $^{125}$Sb, $^{144}$Ce, and $^{194}$Os 
produced in a future supernova at a distance of 10 kpc
would be detectable for a
detector with a sensitivity of
$\sim 10^{-7}\ \gamma~{\rm cm}^{-2}~{\rm s}^{-1}$ at
$E_\gamma\approx 100$--700 keV. 
Note that although $^{144}$Ce may have
decayed substantially when the supernova becomes 
transparent to gamma rays, the
flux at $E_\gamma=134$ keV from its decay may 
still be present at the level of
$\sim 10^{-7}\ \gamma~{\rm cm}^{-2}~{\rm s}^{-1}$.
To accurately compare the relative production of these
nuclei, one should multiply any detected
gamma-ray flux(es) from 
the decay of each nucleus by the
factor $\exp(t/\bar\tau)$, where $t$ is the time between 
detection of neutrinos and detection of the gamma-ray flux(es),
to account for the decay of the nucleus.
If the relative production of these nuclei were indeed specified by the
solar $r$-process abundance pattern, the corrected fluxes would be
proportional to $(X_{\odot}^r/A)(I_\gamma/\bar\tau)$. The distance to the
supernova is not needed for establishing this proportionality.
However, a really accurate determination
may need a much more sophisticated
gamma-ray transport calculation (e.g., Woosley, Pinto, \& Hartmann 1989)
than what we have indicated here.

It is interesting to consider supernova $r$-process events which
produce $r$-process abundance patterns different from that in the
solar system. In fact, distinct supernova sources for the $r$-process
nuclei below and above $A\sim 140$ may be required to explain the meteoritic
data on the $^{129}$I/$^{127}$I and $^{182}$Hf/$^{180}$Hf abundance ratios
in the early solar system (Wasserburg, Busso, \& Gallino 1996). 
For definiteness, we
consider our specific scenario where 
the high-frequency supernova source (case H)
with $f_{\rm SN}^{\rm H}\sim (30\ {\rm yr})^{-1}$ is mainly responsible for
the $r$-process nuclei near and above $A\sim 195$, 
while the low-frequency one
(case L) with $f_{\rm SN}^{\rm L}\sim (300\ {\rm yr})^{-1}$ is mainly
responsible for the $r$-process nuclei near $A\sim 130$ and the bulk of
those between $A\sim 130$ and 195 (Qian, Vogel, \& Wasserburg 1998). 
In this scenario,
only gamma-ray fluxes
from the decay of $^{194}$Os would be detected
at the level of $\sim 10^{-7}\ \gamma~{\rm cm}^{-2}~{\rm s}^{-1}$ from
a frequent supernova of case H, and 
emission from the decay of the lighter nuclei
$^{125}$Sb, $^{137}$Cs, $^{144}$Ce, and $^{155}$Eu would be unobservable 
in this case. 
On the other hand, gamma-ray fluxes
from the decay of $^{125}$Sb, $^{137}$Cs, $^{144}$Ce, and $^{155}$Eu would
all be above $\sim 10^{-7}\ \gamma~{\rm cm}^{-2}~{\rm s}^{-1}$ from
an infrequent supernova of case L, with the expected fluxes $\sim 10$ times
higher than those given in Table 1 where 
a uniform single $r$-process source with $f_{\rm SN}=(30\ {\rm yr})^{-1}$
has been assumed 
to estimate the expected amounts of production $\delta M$.
Since supernovae of cases H and L should eject 
the same amount of mass in the $r$-process nuclei near and above $A\sim 195$
(Qian et al. 1998),
gamma-ray fluxes from the decay
of $^{194}$Os might also be detected at the level given in Table 1 from an
infrequent supernova of case L.

We note that our speculative
association of black hole remnants with supernovae of case H and
neutron star remnants with those of case L
can be tested by pulsar searches 
in addition to observation for
gamma-ray emission from the decay of $r$-process nuclei
after a future Galactic supernova occurs. If our speculation were correct,
only about one out of ten supernovae would leave behind a neutron star. This
would result in many fewer pulsar--supernova-remnant associations than
usually believed.

\subsection{Fluxes from the Vela Supernova Remnant}

Detection of gamma-ray emission from
the decay of $^{126}$Sn in the Vela supernova remnant would prove
that the $r$-process nuclei near $A\sim 126$ are produced
in supernovae. If a gamma-ray detector with a sensitivity of
$\sim 10^{-7}\ \gamma~{\rm cm}^{-2}~{\rm s}^{-1}$ at
$E_\gamma\approx 100$--700 keV can be developed,
this detection is likely to be the first of its kind to connect $r$-process
nucleosynthesis with gamma-ray astronomy since there is no need to wait 
for a new supernova. If it turns out that gamma-ray
emission at $E_\gamma<100$ keV (i.e., hard X-ray emission) is
easier to detect, then the line at $E_\gamma=87.6$ keV 
from the decay of $^{126}$Sn may merit attention (see Table 2). 

We note that the estimates in Table 2 for 
gamma-ray fluxes from the decay of $^{126}$Sn in the Vela supernova remnant 
are subject to uncertainties in the distance to the Vela pulsar and
in the amount of $^{126}$Sn produced in the associated supernova.
The uncertainty in the distance
is basically known since different estimates give $d\approx 125$--500 pc.
The uncertainty in the amount of $^{126}$Sn 
produced in an individual supernova is not known.
As mentioned in \S3.1, meteoritic
data on the $^{129}$I/$^{127}$I and $^{182}$Hf/$^{180}$Hf abundance ratios
in the early solar system seem to require distinct supernova sources for
the $r$-process nuclei below and above $A\sim 140$ (Wasserburg et al.
1996). In our specific scenario, the $r$-process nuclei near $A\sim 130$
are mainly produced in supernovae of case L, which have a frequency of
$f_{\rm SN}^{\rm L}\sim (300\ {\rm yr})^{-1}$ and are expected to leave
behind neutron star remnants (Qian et al. 1998).
In this case, the amount of $^{126}$Sn production $\delta M$ 
in equation (\ref{vela})
would be $\sim 5\times 10^{-6}$ for the
supernova of case L associated with the Vela pulsar (cf. eq. [\ref{mass}]), 
and the resulting 
gamma-ray fluxes from the decay of $^{126}$Sn
would be above $2.2\times 10^{-7}\ \gamma~{\rm cm}^{-2}~{\rm s}^{-1}$ 
at $E_\gamma=415$, 666, and 695 keV even for the largest distance 
$d\approx 500$ pc estimated for this pulsar. Therefore, we are
hopeful that detection of these fluxes will be accomplished 
once a gamma-ray detector with the desirable sensitivity of 
$\sim 10^{-7}\ \gamma~{\rm cm}^{-2}~{\rm s}^{-1}$ at
$E_\gamma\approx 100$--700 keV is developed.

\subsection{Diffuse Fluxes from $^{126}$Sn Decay and NS-NS Mergers}

Many supernovae occur in the Galaxy during the lifetime of
$^{126}$Sn. This would result in the 
diffuse gamma-ray fluxes estimated in Table 2
if supernovae were the main source for
the $r$-process nuclei near $A\sim 126$. 
However, if NS-NS mergers were the main source for these nuclei,
there would be
no diffuse fluxes from the decay of $^{126}$Sn in the Galaxy.
This is because with
an estimated frequency of $\sim (10^6\ {\rm yr})^{-1}$ 
for Galactic NS-NS mergers
(Phinney 1991), the average interval between successive production
events is much longer than the lifetime of $^{126}$Sn. 
We suggest that the design of a gamma-ray
detector with a point-source sensitivity of
$\sim 10^{-7}\ \gamma~{\rm cm}^{-2}~{\rm s}^{-1}$ take into
account the possibility of detecting diffuse fluxes from
$^{126}$Sn decay so that we may determine whether supernovae or 
NS-NS mergers are the main source for the $r$-process nuclei 
near $A\sim 126$. 
We note that if NS-NS mergers were the main source for these nuclei, 
point-source fluxes from $^{126}$Sn decay would be very high from
an event within the last $\sim 10^5$ years
($F_\gamma\sim 2.9\times 10^{-6}\ \gamma~{\rm cm}^{-2}~{\rm s}^{-1}$ 
at $E_\gamma=415$, 666, and 695 keV from an event at 
a distance of $\sim 10$ kpc).

We consider fluxes from $^{126}$Sn decay as
a very promising objective for gamma-ray astronomy related to
$r$-process nucleosynthesis.
Because its lifetime of $1.44\times 10^5$ yr is much longer than the
age of the Vela pulsar, gamma-ray fluxes from the decay of
$^{126}$Sn in the Vela supernova remnant can exist
for a very long time into the future, and
detection of these fluxes can prove that the $r$-process nuclei near
$A\sim 126$ are produced in supernovae. Furthermore, because its lifetime
is much longer than the average interval 
between successive Galactic supernovae
but much shorter than that between successive Galactic NS-NS mergers, 
detection of diffuse gamma-ray fluxes from the decay of $^{126}$Sn
can eliminate NS-NS mergers as the main source for the
$r$-process nuclei near $A\sim 126$. 
Finally, if as in our proposed scenario, only rare supernovae associated
with neutron star remnants were responsible for the $r$-process nuclei near
$A\sim 130$,
then gamma-ray fluxes from the decay of $^{126}$Sn in the Vela supernova
remnant would be much easier to detect.

\section{Conclusion}

We have discussed detection possibilities 
for gamma-ray emission from the decay of $r$-process nuclei
produced in supernovae and their significance for our understanding of
$r$-process nucleosynthesis.
In particular, we have found that a gamma-ray detector
with a sensitivity of $\sim 10^{-7}~\gamma$~cm$^{-2}$~s$^{-1}$
at $E_\gamma\approx 100$--700 keV may detect
the emission from the decay of $^{125}$Sb, $^{137}$Cs, $^{144}$Ce,
$^{155}$Eu, and $^{194}$Os produced in a future Galactic supernova.
In addition, such a detector may detect
the emission from the decay of $^{126}$Sn in the Vela supernova remnant
and the diffuse emission from the decay of $^{126}$Sn produced by
past supernovae in our Galaxy.
The required detector sensitivity is similar to what is projected 
for the proposed detector ATHENA. 
Both the detection of gamma-ray emission from the
decay of several $r$-process nuclei
(e.g., $^{125}$Sb and $^{194}$Os) produced in future
Galactic supernovae and the detection of emission from the decay of
$^{126}$Sn in the Vela supernova remnant would prove that 
supernovae are a site of the $r$-process. Furthermore, the former detection
would allow us to determine whether or not
the $r$-process nuclei are produced in relative proportions specified by
the solar $r$-process abundance pattern in supernova $r$-process events.
Finally, detection of diffuse gamma-ray emission from the decay of 
$^{126}$Sn in our Galaxy would eliminate
neutron-star--neutron-star mergers as the main source for the $r$-process
nuclei near $A\sim 126$. 
In view of these returns, 
we strongly urge that a gamma-ray detector 
with a sensitivity of $\sim 10^{-7}~\gamma$~cm$^{-2}$~s$^{-1}$
at $E_\gamma\approx 100$--700 keV be developed in the near future.

\acknowledgments

We want to thank John Beacom for helping us search the Table of Isotopes,
and Steve Boggs, Fiona Harrison, and Tom Prince
for discussions regarding gamma-ray detectors.
This work was supported in part by the US Department of Energy under
Grant No. DE-FG03-88ER-40397, by NASA under Grant No. NAG5-4076, and 
by Division Contribution No. 8512(992). Y.-Z. Qian was supported by
the David W. Morrisroe Fellowship at Caltech.

\clearpage

\begin{deluxetable}{lcccccc}
\footnotesize
\tablecaption{Expected Gamma-Ray Fluxes from a Galactic Supernova}
\tablehead{
\colhead{Decay Chains of} & \colhead{$X_{\odot}^r$\tablenotemark{a}} & 
\colhead{$\delta M$\tablenotemark{b}} & \colhead{$\bar\tau$\tablenotemark{c}} & 
\colhead{$E_\gamma$} & \colhead{} & \colhead{$F_\gamma$\tablenotemark{d}}\\
\colhead{$r$-Process Nuclei} & \colhead{$(\times 10^{-9})$} &
\colhead{$(10^{-7}\,M_\odot)$} &
\colhead{(yr)} & \colhead{(keV)} & \colhead{$I_\gamma$} &
\colhead{$(10^{-7}\ \gamma~{\rm cm}^{-2}~{\rm s}^{-1})$}
}
\startdata
${^{125}{\rm Sb}}\rightarrow{^{125}{\rm Te}}$ 
& 0.85 & 2.6 & 3.98 & 35.5 & 0.043 & 0.70\nl
& & & & 176 & 0.068 & 1.1\nl
& & & & 380 & 0.015 & 0.25\nl
& & & & 428 & 0.296 & 4.8\nl
& & & & 463 & 0.105 & 1.7\nl 
& & & & 601 & 0.179 & 2.9\nl
& & & & 607 & 0.050 & 0.82\nl
& & & & 636 & 0.113 & 1.8\nl
& & & & 671 & 0.018 & 0.29\nl
${^{137}{\rm Cs}}\rightarrow{^{137}{\rm Ba}}$ 
& 0.26 & 0.77 & 43.4 & 662 & 0.851 & 0.35\nl
${^{144}{\rm Ce}}\rightarrow{^{144}{\rm Pr}}\rightarrow{^{144}{\rm Nd}}$ 
& 0.44 & 1.3 & 1.12 & 80.1 & 0.014 & 0.35\nl
& & & & 134 & 0.111 & 2.9\nl
& & & & 697 & 0.013 & 0.35\nl
& & & & 2186 & 0.007 & 0.18\nl
${^{155}{\rm Eu}}\rightarrow{^{155}{\rm Gd}}$ 
& 0.18 & 0.54 & 6.87 & 86.5 & 0.307 & 0.50\nl
& & & & 105 & 0.212 & 0.34\nl
${^{194}{\rm Os}}\rightarrow{^{194}{\rm Ir}}\rightarrow{^{194}{\rm Pt}}$
& 2.1 & 6.4 & 8.66 & 43.1 & 0.054 & 0.65\nl
& & & & 294 & 0.026 & 0.31\nl
& & & & 328 & 0.131 & 1.6\nl 
& & & & 645 & 0.012 & 0.14\nl
\enddata
\tablenotetext{a}{Solar $r$-process mass fractions of the stable
nuclei in the decay chains.}
\tablenotetext{b}{Amounts of production per supernova for 
the $r$-process nuclei at the beginning of the decay chains estimated
from equation (\protect{\ref{mass}}) with $M_G=10^{11}\,M_\odot$,
$t_G=10^{10}$ yr, and $f_{\rm SN}=(30\ {\rm yr})^{-1}$.}
\tablenotetext{c}{Lifetimes of the $r$-process nuclei 
at the beginning of the decay chains.}
\tablenotetext{d}{Gamma-ray fluxes 
estimated from equation (\protect{\ref{flux}})
for a supernova at a distance of 10 kpc.}
\end{deluxetable}

\clearpage

\begin{deluxetable}{lccc}
\footnotesize
\tablecaption{Expected Gamma-Ray Fluxes from the Decay of 
$^{126}$Sn\tablenotemark{a}}
\tablehead{
\colhead{$E_\gamma$} & \colhead{} & 
\colhead{$F_\gamma$~(Vela)\tablenotemark{b}} & 
\colhead{${\cal{F}}_\gamma$~(diffuse)\tablenotemark{c}}\\
\colhead{(keV)} & \colhead{$I_\gamma$} & 
\colhead{$(10^{-7}\ \gamma~{\rm cm}^{-2}~{\rm s}^{-1})$} &
\colhead{$(10^{-7}\ \gamma~{\rm cm}^{-2}~{\rm s}^{-1}~{\rm rad}^{-1})$}
}
\startdata
23.3 & 0.064 & 0.14 & 0.32\nl
64.3 & 0.096 & 0.21 & 0.48\nl
86.9 & 0.089 & 0.19 & 0.44\nl
87.6 & 0.370 & 0.80 & 1.8\nl
415 & 0.976 & 2.1 & 4.9\nl
666 & 0.999 & 2.2 & 5.0\nl
695 & 0.965 & 2.1 & 4.8\nl
721 & 0.075 & 0.16 & 0.37\nl
\enddata
\tablenotetext{a}{The lifetime of $^{126}$Sn is $1.14\times 10^5$ yr.
The decay chain is 
${^{126}{\rm Sn}}\rightarrow{^{126}{\rm Sb}}\rightarrow{^{126}{\rm Te}}$.}
\tablenotetext{b}{Gamma-ray fluxes 
from the Vela supernova remnant (taken to be at a distance of 200 pc and
having $5.0\times 10^{-7}\,M_\odot$ of $^{126}$Sn)
(see eq. [\protect{\ref{vela}}]).}
\tablenotetext{c}{Diffuse gamma-ray fluxes 
in the Galactic center direction
for a uniform $^{126}$Sn 
production rate of $p=1.65\times 10^{-8}\,M_\odot~{\rm yr}^{-1}$
in the Galaxy
(see eq. [\protect{\ref{diff}}]).}

\end{deluxetable}

\end{document}